\documentclass[12pt,a4paper]{article}
\usepackage{amsmath}
\DeclareMathOperator{\EOM}{EOM}
\allowdisplaybreaks

{\def\$#1: #2 ${#2}}
\begin{document}

\title{%
  Heavy Quark Effective Field Theory
    at~$\mathcal{O}(1/m_Q^2)$. I.\\
  QCD Corrections to the Lagrangian}

\author{%
  Christopher Balzereit\thanks{%
      Supported by the German National Scholarship Foundation.}%
    \thanks{email:
      \texttt{chb@crunch.ikp.physik.th-darmstadt.de}} \\
  Thorsten Ohl\thanks{e-mail:
      \texttt{Thorsten.Ohl@Physik.TH-Darmstadt.de}} \\
  \hfil \\
  Technische Hochschule Darmstadt \\
  Schlo\ss gartenstr. 9 \\
  D-64289 Darmstadt \\
  Germany}

\date{%
  IKDA 96/11 \\
  hep-ph/9604352 \\
  April 1996 \\\hfil}

%%%%%%%%%%%%%%%%%%%%%%%%%%%%%%%%%%%%%%%%%%%%%%%%%%%%%%%%%%%%%%%%%%%%%%%%
\maketitle
\begin{abstract}
  We present a new calculation of the renormalized HQET Lagrangian at
  order~$\mathcal{O}(1/m_Q^2)$ and discuss the consequences of the BRST
  invariance of QCD and the reparameterization invariance of HQET.
  Our result corrects earlier, conflicting calculations and sets the
  stage for the calculation of the renormalized currents at
  order~$\mathcal{O}(1/m_Q^2)$.
\end{abstract}
\newpage

%%%%%%%%%%%%%%%%%%%%%%%%%%%%%%%%%%%%%%%%%%%%%%%%%%%%%%%%%%%%%%%%%%%%%%%%
\section{Introduction}
\label{sec:introduction}
Heavy Quark Effective Field Theory
(HQET)~\cite{HQET_classic} has been
established as the theoretical tool of choice for the description of
mesons and baryons containing heavy
quarks~\cite{HQET_review}.  This derives from the fact that it is a
\emph{systematic} expansion in inverse powers of the heavy quark with
well defined and calculable coefficients.  Furthermore, its
realization of the spin and flavor symmetry of the low energy theory
is a phenomenologically powerful tool. 

However, since the expansion parameter~$\Lambda_H/m_Q$ is about~$0.12$
(using a hadronic scale~$\Lambda_H\approx 600\text{MeV}$
and~$m_Q=m_b$) leading order calculations are not sufficient for
precision calculations.  Terms of order~$\mathcal{O}(1/m_Q^2)$
\emph{including} leading QCD corrections have to be under control.

An indispensable prolegomenon to the calculation of renormalized
matrix elements of currents is the renormalization of the Lagrangian.
Unfortunately, two calculations with conflicting
results~\cite{Lee:1991:HQET, Balk/etal:1993:HQET} have been reported.
In this note we present the result of a new calculation of the
renormalized Lagrangian, which differs from the previous two.  We will
demonstrate that our result satisfies important consistency conditions
that are violated by the earlier calculations.

This note is organized as follows: in section~\ref{sec:basis} we introduce
our operator basis.  Our result for the anomalous dimensions is presented
in section~\ref{sec:anodim}.  In section~\ref{sec:consistency} we will
discuss the consistency of this result and compare it in
section~\ref{sec:comparison} with earlier calculations.  Finally, we
present the renormalization group flow in section~\ref{sec:RG-flow} and
conclude in section~\ref{sec:conclusions}.

The renormalized currents will be presented in a subsequent
note~\cite{Balzereit/Ohl:1996:currents} and phenomenological applications
will appear later~\cite{Balzereit/Ohl:1996:V_cb}.  A more detailed
discussion of technical matters will be presented
in~\cite{Balzereit/Ohl:1996:technical}.

%%%%%%%%%%%%%%%%%%%%%%%%%%%%%%%%%%%%%%%%%%%%%%%%%%%%%%%%%%%%%%%%%%%%%%%%
\section{Operator basis}
\label{sec:basis}
The Lagrangian of HQET is defined by a systematic expansion of QCD in
inverse powers of the heavy quark mass
\begin{equation} 
  \mathcal{L}_{\text{HQET}} = \bar h_{v}(i v D) h_{v}
    + \frac{1}{2m_Q} \sum_i \breve C_i^{(1)} \mathcal{O}_i^{(1)}
  \mbox{} + \frac{1}{(2m_Q)^2} \sum_i \breve C_i^{(2)} \mathcal{O}_i^{(2)}
    + \mathcal{O}(\frac{1}{(2m_Q)^3}) \,.
\end{equation}
At order~$\mathcal{O}(1/m_Q^0)$ there is only one
operator~$\bar h_v(ivD)h_v$, which is independent of the spin and 
flavor of the quark, resulting in the celebrated spin-flavor symmetry
of HQET.

At order~$\mathcal{O}(1/m_Q)$ there are three independent operators.
We shall use the conventional basis
\begin{subequations}
\begin{align}
  \mathcal{O}^{(1)}_1
    &= \bar h_v (iD)^2 h_v \\
  \mathcal{O}^{(1)}_2
    &= \frac{g}{2} \bar h_v \sigma^{\mu \lambda} 
       F_{\mu \lambda} h_v \\
  \mathcal{O}^{(1)}_3
    &= \bar h_v (ivD)^2 h_v \,.
\end{align}
\end{subequations}
Below, the operators~$\mathcal{O}^{(1)}_1$ and~$\mathcal{O}^{(1)}_2$
will also be called \emph{kinetic} and \emph{chromo-magnetic}
respectively.  The operator~$\mathcal{O}^{(1)}_3$ vanishes by the
equations of motion (EOM)
\begin{equation}
  ivDh_v |Q\rangle = 0
\end{equation}
for heavy quark states~$|Q\rangle$.  For the renormalization of the
Lagrangian, it is not necessary to include this operator, if the EOM
are used consistently.  It is however needed as a counterterm in the
renormalization of the heavy quark
currents~\cite{Balzereit/Ohl:1996:currents} and will be included here.
In addition, inclusion of the operators vanishing by the EOM allows to
make the relations following from reparameterization invariance
explicit. Finally, the extraction of coefficients with the aid of
symbolic manipulation programs is more straightforward in the full
basis, while the work induced by the additional operators is
insignificant.

At order~$\mathcal{O}(1/m_Q^2)$ there are thirteen independent
operators. They are grouped in four classes.  Two of the local
operators do not vanish by the EOM.  We will denote them collectively
by~$\vec\mathcal{O}^{(2)}$ and choose them as
\begin{subequations}
\begin{align}
  \mathcal{O}^{(2)}_1
    &= \bar h_v iD_{\mu} (ivD) iD^{\mu} h_v \\
  \mathcal{O}^{(2)}_2
    &= \bar h_v i \sigma^{\mu \lambda} iD_{\mu}
       (ivD) iD_{\lambda} h_v\,. \\
\intertext{%
    The five remaining local operators vanish by the EOM.  We denote
    them by $\vec\mathcal{O}^{(2)}_{\EOM}$ and choose the basis}
  \mathcal{O}^{(2)}_3
    &= \bar h_v (ivD) (iD)^2 h_v \\
  \mathcal{O}^{(2)}_4
    &= \bar h_v (iD)^2 (ivD) h_v  \\
  \mathcal{O}^{(2)}_5
    &= \bar h_v (ivD)^3 h_v \\
  \mathcal{O}^{(2)}_6
    &= -\frac{g}{2} \bar h_v (ivD)  \sigma^{\mu \lambda} 
       F_{\mu \lambda} h_v  \\
  \mathcal{O}^{(2)}_7
    &= -\frac{g}{2} \bar h_v  \sigma^{\mu \lambda}
       F_{\mu \lambda} (ivD) h_v \,. \\
\intertext{%
    In addition to the local operators, there are the time-ordered
    products of the lower dimensional operators.  There are three of them
    that do not vanish by the EOM.  They will be
    denoted~$\vec\mathcal{T}^{(2)}$.}
  \mathcal{T}^{(2)}_{11}
    &= \frac{i}{2} T \left\{ \left[\bar h_{v}(iD)^{2}h_{v}\right]
                             \left[\bar h_{v}(iD)^{2}h_{v}\right] \right\} \\
  \mathcal{T}^{(2)}_{12}
    &= \frac{ig}{2} T \left\{ \left[\bar h_{v}(iD)^{2}h_{v}\right]
                              \left[\bar h_{v} \sigma^{\mu \lambda} 
                                    F_{\mu \lambda} h_{v}\right] \right\} \\
  \mathcal{T}^{(2)}_{22}
    &= \frac{ig^2}{8} T \left\{ \left[\bar h_{v} \sigma^{\mu \lambda}
                                      F_{\mu \lambda} h_{v}\right]
                                \left[\bar h_{v} \sigma^{\mu \lambda} 
                                      F_{\mu \lambda} h_{v}\right] \right\} \\
\intertext{%
    Below, the operators~$\mathcal{T}^{(2)}_{11}$
    and~$\mathcal{T}^{(2)}_{22}$ will also be called
    \emph{double-kinetic} and \emph{double-chromo-magnetic} respectively.
    Finally there are three more
    time-ordered products~$\vec\mathcal{T}^{(2)}_{\EOM}$ that vanish
    by the EOM}
  \mathcal{T}^{(2)}_{13}
    &= i\, T \left\{ \left[\bar h_{v}(iD)^{2}h_{v}\right]
                     \left[\bar h_{v}(ivD)^{2}h_{v}\right] \right\} \\
  \mathcal{T}^{(2)}_{23}
    &= \frac{ig}{2} T \left\{ \left[\bar h_{v} \sigma^{\mu \lambda} 
                                    F_{\mu \lambda} h_{v}\right]
                              \left[\bar h_{v}(ivD)^{2}h_{v}\right] \right\}\\
  \mathcal{T}^{(2)}_{33}
    &= \frac{i}{2} T \left\{ \left[\bar h_{v}(ivD)^{2}h_{v}\right]
                             \left[\bar h_{v}(ivD)^{2}h_{v}\right] \right\}\,.
\end{align}
\end{subequations}
Below we shall refer to the operators vanishing by the equations of
motion as \emph{EOM operators}, for short.

For calculational convenience, we have not chosen a manifestly
hermitian basis, but the results presented below show that indeed only
hermitian linear combinations show up in counter terms.

%%%%%%%%%%%%%%%%%%%%%%%%%%%%%%%%%%%%%%%%%%%%%%%%%%%%%%%%%%%%%%%%%%%%%%%%
\section{Anomalous dimensions}
\label{sec:anodim}
The most convenient approach to the calculation of anomalous
dimensions uses the background field gauge~\cite{BFM}
with gauge fixing term
\begin{equation}
\label{eq:gauge-fix}
  -\frac{1}{2\xi} \left( D_\mu(V)A^\mu \right)^2 \,,
\end{equation}
because only the renormalization constants of those operators that
are \emph{manifestly} invariant under gauge transformations of the
background field have to be calculated.  Therefore, only the
divergent three-point functions have to be calculated to derive the
anomalous dimensions.  Furthermore, the Ward identities for the
classical background fields are particularly simple and provide a
powerful tool for checking our results.

The background field gauge with an arbitrary gauge parameter allows an
independent test of the consistency of our results by comparing
the~$\xi$ dependence with general
results~\cite{Kluberg-Stern/Zuber:1975:BFM/operators} for the one-loop
effective action.  Below we shall write~$\bar\xi=1-\xi$ for the gauge
parameter.

The anomalous dimensions at order~$\mathcal{O}(1/m_Q)$ are well
known~\cite{HQET_renormalization}
(the gauge parameter dependence
has been calculated in~\cite{Kilian:1994:thesis}):
\begin{equation}
\label{eq:anodim1}
  \hat\gamma^{(1)} =
    \begin{pmatrix}
      0 & 0               & 2C_F^\xi \\
      0 & -\frac{1}{2}C_A & 0        \\
      0 & 0               & -C_F^\xi
  \end{pmatrix}\,,
\end{equation}
where we have introduced the shorthand~$C_F^\xi = C_F(1+\bar\xi/2)$.
Here and below, we have extracted the common loop factor~$\alpha/\pi$
from~(\ref{eq:anodim1}).

The matrix of anomalous dimensions can naturally be written in block
form, separating local operators from non-local operators and
separating EOM operators from the rest:
\begin{equation}
  \hat\gamma^{(2)} =
    \begin{matrix}
      \vphantom{{\displaystyle\int}}
        & \begin{matrix}
              \vec{\mathcal{O}}^{(2)}
            & \vec{\mathcal{O}}^{(2)}_{\EOM}
            & \vec{\mathcal{T}}^{(2)}
            & \vec{\mathcal{T}}^{(2)}_{\EOM}
          \end{matrix} \\
      \begin{matrix}
        \vec{\mathcal{O}}^{(2)}\\
        \vec{\mathcal{O}}^{(2)}_{\EOM}\\
        \vec{\mathcal{T}}^{(2)}\\
        \vec{\mathcal{T}}^{(2)}_{\EOM}
      \end{matrix} &
      \begin{pmatrix}
        \begin{matrix}
          \hat\gamma^{(2)}_{\mathrm{l}}
            & \hat\gamma^{(2)\EOM}_{\mathrm{l},1} \\
          0 & \hat\gamma^{(2)\EOM}_{\mathrm{l},2}
        \end{matrix} & \mbox{\Large$0$} \\
        \begin{matrix}
          \hat\gamma^{(2)}_{\mathrm{m}}
            & \hat\gamma^{(2)\EOM}_{\mathrm{m},1} \\
          0 & \hat\gamma^{(2)\EOM}_{\mathrm{m},2}
        \end{matrix} &
        \begin{matrix}
          \hat\gamma^{(2)}_{\mathrm{n}}
            & \hat\gamma^{(2)\EOM}_{\mathrm{n},1} \\
          0 & \hat\gamma^{(2)\EOM}_{\mathrm{n},2}
        \end{matrix}
      \end{pmatrix}
    \end{matrix} \,.
\end{equation}
The upper right block has to vanish, because Weinberg's
theorem~\cite{Weinberg:1960:theorem} guarantees that the
renormalization of the local operators does not require counterterms
from the time-ordered products.  The other three blocks display the
block triangular structure required by the fact that the
renormalization of EOM operators only induces counterterms that are 
EOM operators themselves.

The anomalous dimensions have been calculated manually.  These
calculations have been verified with the help
of~\texttt{FORM}~\cite{Vermaseren:1991:FORM2} as a warm-up for the
renormalization of the currents~\cite{Balzereit/Ohl:1996:currents},
which requires the use of symbolic manipulation programs for
economical reasons.

We start the presentation of the results with the renormalization of
the local operators
\begin{subequations}
\label{eq:anodim2}
\newcommand{\dto}{\Leftarrow}
\begin{align}
  \hat\gamma^{(2)}_{\mathrm{l}}
    &= \begin{pmatrix}
         -\frac{1}{3}C_A & 0  \\
                       0 & 0
       \end{pmatrix} \\
  \hat\gamma^{(2)\EOM}_{\mathrm{l},1}
    &= \begin{pmatrix}
         \frac{1}{6}C_A & \frac{1}{6}C_A & -2C_F(1+\bar\xi)
           & 0               & 0       \\
         0              & 0              & 0
           & -\frac{3}{4}C_A & -\frac{3}{4}C_A
       \end{pmatrix} \\
  \hat\gamma^{(2)\EOM}_{\mathrm{l},2}
    &= \begin{pmatrix}
         \begin{matrix}
           0 & 0 & -C_F(1+2\bar\xi)  \\
           0 & 0 & -C_F(1+2\bar\xi)  \\
           0 & 0 & -C_F^\xi
         \end{matrix} & \mbox{\Large$0$} \\
           \mbox{\Large$0$}
             & \begin{matrix}
                 -C_A & 0   \\
                 0    & -C_A
               \end{matrix}
       \end{pmatrix} \,. \\
\intertext{%
    The renormalization of the time-ordered products requires local
    counterterms as well.  The corresponding anomalous dimensions are}
\label{eq:anodim2-mix}
  \hat\gamma^{(2)}_{\mathrm{m}}
    &= \begin{pmatrix}
         -\frac{1}{6}C_A - \frac{8}{3}C_F & 0  \\
         0                                & -C_A \\
         -\frac{5}{6}C_A                  & 0
       \end{pmatrix} \\
  \hat\gamma^{(2)\EOM}_{\mathrm{m},1}
    &= \begin{pmatrix}
          C_{AF}^\xi      & C_{AF}^\xi      & -8C_F^\xi
            & 0               & 0               \\
          0               & 0               & 0
            & C_A-2C_F^\xi    & C_A-2C_F^\xi    \\
          \frac{5}{12}C_A & \frac{5}{12}C_A & 2C_F
            & -\frac{3}{4}C_A & -\frac{3}{4}C_A
       \end{pmatrix} \\
\intertext{%
    using the
    shorthand~$C_{AF}^\xi = \frac{1}{12}C_A + C_F(\frac{10}{3}+\bar\xi)$.}
  \hat\gamma^{(2)\EOM}_{\mathrm{m},2}
    &= \begin{pmatrix}
         -C_F^\xi & -C_F^\xi & 6C_F(1+\bar\xi)
           & 0                       & 0 \\
         0        & 0        & 0
           & -\frac{1}{2}C_A+C_F^\xi & -\frac{1}{2}C_A+C_F^\xi \\
         0        & 0        & -C_F^\xi
           & 0                       & 0
        \end{pmatrix}
\end{align}
\end{subequations}
The blocks~$\hat\gamma^{(2)}_{\mathrm{n}}$,
$\hat\gamma^{(2)\EOM}_{\mathrm{n},1}$
and~$\hat\gamma^{(2)\EOM}_{\mathrm{n},2}$ are given by the sum of the
appropriate anomalous dimensions from~$\hat\gamma^{(1)}$.

%%%%%%%%%%%%%%%%%%%%%%%%%%%%%%%%%%%%%%%%%%%%%%%%%%%%%%%%%%%%%%%%%%%%%%%%
\section{Consistency of the results}
\label{sec:consistency}
The symmetries of HQET entail relations among the anomalous
dimensions that can be used to check the result~(\ref{eq:anodim2}).
Such consistency checks are useful in the present case and are of
vital importance in the considerably more involved renormalization of
the HQET currents~\cite{Balzereit/Ohl:1996:currents}.

%%%%%%%%%%%%%%%%%%%%%%%%%%%%%%%%%%%%%%%%%%%%%%%%%%%%%%%%%%%%%%%%%%%%%%%%
\subsection{Gauge invariance}
\label{sec:gauge-invariance}
As alluded to in the previous section, we have used the Ward
identities for the background fields and the $\xi$-dependence of the
one-loop effective action for verifying our results.

From the simple QED-like Ward identity for the background field
\begin{equation}
 q_\alpha\tilde\Gamma^\alpha(p,q) = \tilde S(p+q) - \tilde S(p),
\end{equation}
with~$\tilde S$ and~$\tilde\Gamma^\alpha$ denoting the
one-particle-irreducible two- and three-point functions with one
operator insertion respectively, follows that the counterterms
proportional to~$C_A$ (i.e.~the non-abelian contributions) have to be
transversal.  Our result~(\ref{eq:anodim2}) passes this consistency
check. The technical details will be presented
in~\cite{Balzereit/Ohl:1996:technical}.

In background field gauge, the $\xi$-dependence of the renormalized
effective action at \emph{one loop} order is known to have the
following form~\cite{Kluberg-Stern/Zuber:1975:BFM/operators}
\begin{equation}
\label{eq:xi-dep}
  -2\xi \frac{\partial}{\partial\xi} \tilde\Gamma^\alpha
    = \hat\Gamma^{(0)}_{A^\alpha Q}
        \ast \tilde\Gamma^{(1)}_{J L \bar h_{v} h_{v}}
      + \tilde\Gamma^{(0)}_{\bar h_v A^\alpha h_v}
          \ast \Gamma^{(1)}_{\bar{M} L h_v}
      + \Gamma^{(0)}_{\bar h_v A^\alpha h_v}
          \ast \tilde\Gamma^{(1)}_{\bar M L h_v}
      + \ldots \,
\end{equation}
where~$\Gamma^{(n)}$ denotes the effective action at $n$-loop order, 
$\hat\Gamma^{(n)}$ the same effective action with the gauge fixing
term~(\ref{eq:gauge-fix}) subtracted and~$\tilde\Gamma^{(n)}$ the
effective action with an operator inserted.  Finally, subscripts
denote functional differentiation and~$\ast$ integration over the
corresponding space-time argument.

With the help of power counting we can identify the possible
contributions to the right hand side of~(\ref{eq:xi-dep}) that have
the correct tensor structure.  It can be
shown~\cite{Balzereit/Ohl:1996:technical} for all operators, with the
exception of~$\mathcal{O}^{(2)}_6$ and~$\mathcal{O}^{(2)}_7$, that
such contributions 
have to be proportional to~$C_F$.  The explicit calculation
shows that this feature remains true for~$\mathcal{O}^{(2)}_6$
and~$\mathcal{O}^{(2)}_7$ as well.  Furthermore, there are no
$\xi$-dependent counter terms that do not vanish by the EOM.  Our
result~(\ref{eq:anodim2}) passes these consistency checks as well.

%%%%%%%%%%%%%%%%%%%%%%%%%%%%%%%%%%%%%%%%%%%%%%%%%%%%%%%%%%%%%%%%%%%%%%%%
\subsection{Reparameterization invariance}
\label{sec:reparam-invariance}
The HQET Lagrangian is a reparameterized form of the QCD
Lagrangian, therefore the matching  coefficients~$\breve C$ of different
orders in~$1/m_Q$ are
related~\cite{Luke/Manohar:1992:HQET_reparameterization}.  For example
the matching coefficient of the chromo-magnetic operators
in~$\mathcal{O}(1/m_Q^2)$ can be derived from the coefficient
in~$\mathcal{O}(1/m_Q)$
\begin{equation}
  \breve C^{(2)}_2 = 2 \breve C^{(1)}_2 - 1\,.
\end{equation}

On the other hand, we know that the product of the matching
coefficients and the renormalization constants~$\vec{\breve C}
\hat Z_{\text{MS}}^{-1}$ is finite and we can derive relations between
the~$\mathcal{O}(\alpha)$ matching coefficients, the tree-level
matching coefficients and the anomalous dimensions:
\begin{equation}
\label{eq:reparam}
  \vec{\breve C}^{(\alpha)}
    +\vec{\breve C}^{(\text{tree})} \hat\gamma^{(2)} = 0 \,.
\end{equation}
Since there are reparameterization invariance relations among
the~$\mathcal{O}(\alpha)$ matching coefficients,
(\ref{eq:reparam})~induces reparameterization invariance relations
among the anomalous dimensions~\cite{Balzereit/Ohl:1996:technical}.
These relations are satisfied by our result~(\ref{eq:anodim2}).

%%%%%%%%%%%%%%%%%%%%%%%%%%%%%%%%%%%%%%%%%%%%%%%%%%%%%%%%%%%%%%%%%%%%%%%%
\section{Comparison with earlier calculations}
\label{sec:comparison}
Two calculations of the renormalized HQET Lagrangian at
order~$\mathcal{O}(1/m_Q^2)$ have been circulated as preprints in the
past~\cite{Lee:1991:HQET, Balk/etal:1993:HQET}.  Their results are
not consistent with each other and our result differs from
\emph{both}.  Therefore a brief discussion of the errors in these
calculations seems to be in order:
\begin{itemize}
  \item After transforming the result of~\cite{Lee:1991:HQET} to our
    basis, it turns out that the
    coefficients~$\hat\gamma^{(2)}_{\mathrm{m}}$ of the local counter
    terms for the double insertions are incorrect.  In particular,
    the~$-C_A$ entry in~(\ref{eq:anodim2-mix}) is fixed by
    reparameterization invariance, which is therefore violated by the
    result in~\cite{Lee:1991:HQET}.  The argument is unfortunately
    technically involved and will be presented
    elsewhere~\cite{Balzereit/Ohl:1996:technical}.
  \item The operator basis used in~\cite{Balk/etal:1993:HQET} is
    inconsistent. While these authors have used the EOM in their
    calculations, they do include an operator
    \begin{equation}
       \frac{ig}{2} \left( \bar h_v \sigma^{\alpha\nu} T^a h_v \right)
            D_\alpha F^a_{\mu\nu} v^\mu
       = -\frac{g}{4} \bar h_v \sigma^{\mu\nu}
            \left[ F_{\mu\nu}, ivD \right] h_v
    \end{equation}
    in their basis, which vanishes by this equation of motion.  On the
    other hand, they have missed a spin-symmetric operator that is
    required as a counter term for the double insertions of the
    kinetic and chromo-magnetic operators.
\end{itemize}

%%%%%%%%%%%%%%%%%%%%%%%%%%%%%%%%%%%%%%%%%%%%%%%%%%%%%%%%%%%%%%%%%%%%%%%%
\section{Renormalization group flow}
\label{sec:RG-flow}

An analytical solution of the renormalization group equation
using~$(\ref{eq:anodim2})$ in the full basis seems to be
impracticable.  We can however restrict ourselves to the basis of the
operators~$\vec{\mathcal{O}}^{(2)}\vec{\mathcal{T}}^{(2)}$ that do not
vanish by the EOM:
\begin{equation}
  \hat\gamma_{\text{phys}}^{(2)} =
    \begin{pmatrix}
      -\frac{1}{3}C_A                  & 0    & 0 & 0               & 0    \\
      0                                & 0    & 0 & 0               & 0    \\
      -\frac{1}{6}C_A - \frac{8}{3}C_F & 0    & 0 & 0               & 0    \\
      0                                & -C_A & 0 & -\frac{1}{2}C_A & 0    \\
      -\frac{5}{6}C_A                  & 0    & 0 & 0               & -C_A
    \end{pmatrix}\,.
\end{equation}
%%% \begin{dubious}
%%%   I would prefer if it looked like this, but with the first two
%%%   columns lined up:
%%%   \begin{equation*}
%%%     \hat\gamma_{\text{phys}}^{(2)} =
%%%       \begin{pmatrix}
%%%         \begin{matrix}
%%%           -\frac{1}{3}C_A                  & 0 \\
%%%           0                                & 0 \\
%%%           -\frac{1}{6}C_A - \frac{8}{3}C_F & 0
%%%         \end{matrix} & \mbox{\Large$0$} \\
%%%         \begin{matrix}
%%%           0               & -C_A \\
%%%           -\frac{5}{6}C_A & 0
%%%         \end{matrix} &
%%%         \begin{matrix}
%%%           0 & -\frac{1}{2}C_A & 0 \\
%%%           0 & 0               & -C_A
%%%         \end{matrix}
%%%       \end{pmatrix}
%%%   \end{equation*}
%%% \end{dubious}
This reduced renormalization group equation
\begin{equation}
  \frac{d}{d\ln\mu}\vec C^{(2)}(\mu)
    + \frac{\alpha}{\pi}
        \hat\gamma_{\text{phys}}^{(2)\top}\vec{C}^{(2)}(\mu) = 0
\end{equation}
with the initial (matching) conditions
\begin{subequations}
\begin{align}
  C^{(2)}_1(m_Q) &= -1 \\
  C^{(2)}_2(m_Q) &= 1 \\
  C^{(2)}_{11}(m_Q) &= C^{(1)}_1(m_Q) C^{(1)}_1(m_Q) = 1 \\
  C^{(2)}_{12}(m_Q) &= C^{(1)}_1(m_Q) C^{(1)}_2(m_Q) = 1 \\
  C^{(2)}_{22}(m_Q) &= C^{(1)}_2(m_Q) C^{(1)}_2(m_Q) = 1
\end{align}
\end{subequations}
can be solved analytically
\begin{subequations}
\label{eq:RGE:solution}
\begin{align}
  C^{(2)}_1(\mu)
    &= \left(\frac{8C_F}{C_A} - \frac{7}{4}\right)
         \left(\frac{\alpha(\mu)}{\alpha(m_Q)}
           \right)^{\displaystyle -\frac{C_A}{6\beta^{(1)}}}
       + \frac{5}{4}
         \left(\frac{\alpha(\mu)}{\alpha(m_Q)}
           \right)^{\displaystyle -\frac{C_A}{2\beta^{(1)}}}
       - \frac{8C_F}{C_A} - \frac{1}{2}  \\
  C^{(2)}_2(\mu)
    &= 2\left(\frac{\alpha(\mu)}{\alpha(m_Q)}
          \right)^{\displaystyle -\frac{C_A}{4\beta^{(1)}}} - 1 \\
  C^{(2)}_{11}(\mu)
    &= C^{(1)}_1(\mu) C^{(1)}_1(\mu) = 1 \\
  C^{(2)}_{12}(\mu)
    &= C^{(1)}_1(\mu) C^{(1)}_2(\mu)
       = \left(\frac{\alpha(\mu)}{\alpha(m_Q)}
           \right)^{\displaystyle -\frac{C_A}{4\beta^{(1)}}} \\
  C^{(2)}_{22}(\mu)
    &= C^{(1)}_2(\mu) C^{(1)}_2(\mu)
       = \left(\frac{\alpha(\mu)}{\alpha(m_Q)}
           \right)^{\displaystyle -\frac{C_A}{2\beta^{(1)}}} \,.
\end{align}
\end{subequations}
Flavor threshold at which~$\beta^{(1)}$ changes have been ignored
in~(\ref{eq:RGE:solution}).  It is straightforward to recover them by
pasting solutions together at the thresholds.

The solution of the renormalization group equation in the full basis
can be obtained numerically for specific values of the gauge
parameter~$\xi$ and a specific gauge group (i.e.~$\text{SU}(3)$).
This will be done for the renormalization of the
currents~\cite{Balzereit/Ohl:1996:currents}.

%%%%%%%%%%%%%%%%%%%%%%%%%%%%%%%%%%%%%%%%%%%%%%%%%%%%%%%%%%%%%%%%%%%%%%%%
\section{Conclusions}
\label{sec:conclusions}
We have presented a new calculation of the renormalized HQET
Lagrangian at order~$\mathcal{O}(1/m_Q^2)$.  Our result corrects
previous calculations and obeys the Ward identities imposed by the
BRST invariance of QCD and the reparameterization invariance of HQET.

The new renormalized Lagrangian has been used in a calculation of the
renormalized HQET currents at order~$\mathcal{O}(1/m_Q^2)$.  The
results will be published in a
sequel~\cite{Balzereit/Ohl:1996:currents} to this note.  A more
detailed discussion of the consistency checks provided by BRST and
reparameterization invariance will be presented
elsewhere~\cite{Balzereit/Ohl:1996:technical}, together with technical
details of the calculation of renormalized Lagrangian and currents.

%%%%%%%%%%%%%%%%%%%%%%%%%%%%%%%%%%%%%%%%%%%%%%%%%%%%%%%%%%%%%%%%%%%%%%%%

\end{document}